\documentclass[a4paper,11pt]{article}
\pdfoutput=1 

\usepackage{jcappub} 

\usepackage{subcaption}
\usepackage{graphicx}                
\usepackage{multirow}
\usepackage[T1]{fontenc} 
\usepackage[table]{xcolor}
\usepackage{float}
\def\noi{{\noindent}}

\definecolor{verde}{rgb}{0,0.5,0}

\def\be{\begin{equation}}
\def\ee{\end{equation}}
\def\bea{\begin{eqnarray}}
\def\eea{\end{eqnarray}}
\def\be{\begin{equation}}
\def\ee{\end{equation}}
\def\ba{\begin{eqnarray}}
\def\ea{\end{eqnarray}}
\def\p{\partial}

\title{\boldmath Interferometer Constraints on\\ the Inflationary Field Content}

\author{Laura Iacconi$^{a}$, Matteo Fasiello$^{a}$, Hooshyar Assadullahi$^{a}$, Emanuela Dimastrogiovanni$^{b}$, and David Wands$^{a}$}
\affiliation{$^{a}$Institute of Cosmology \& Gravitation, University of Portsmouth, PO1 3FX, UK}
\affiliation{$^{b}$School of Physics, The University of New South Wales, Sydney, NSW 2052, Australia}

\abstract{ With an energy scale that can be as high as $10^{14}\,{\rm GeV}$, inflation may provide a unique probe of high-energy physics. Both scalar and tensor fluctuations generated during this early accelerated expansion contain crucial information about the particle content of the primordial universe. The advent of ground- and space-based interferometers enables us to probe primordial physics at length-scales much smaller than current CMB constraints. One key prediction of single-field slow-roll inflation is a red-tilted gravitational wave spectrum, currently inaccessible at interferometer scales. Interferometers probe directly inflationary physics that deviates from the minimal scenario and in particular additional particle content with sizeable couplings to the inflaton field. We adopt here an effective description for such fields and focus on the case of extra spin-2 fields. We find that time-dependent sound speeds for the helicity-2 modes can generate primordial gravitational waves with a blue-tilted spectrum, potentially detectable at interferometer scales.}

\begin{document}
	\maketitle
	\flushbottom
	
	\section{Introduction}
	
	\noi The resolution of a number of early puzzles in the ``hot big-bang'' model has secured a central role for inflation in the cosmological standard model. An early phase of accelerated expansion accounts for the nearly uniform temperature observed in the cosmic microwave background (CMB) and provides the seeds for structure formation. The simplest implementation of the inflationary mechanism consists of a scalar field slowly rolling down its potential to produce a sufficiently long expansion. Measurements of the CMB radiation constrain primordial scalar fluctuations around the inflating FLRW background to be nearly Gaussian and almost scale-invariant (at large scales). Pending a detection of the primordial tensor signal, upper limits exist on the gravitational waves spectrum, including a tensor-to-scalar ratio of $r<0.056$ \cite{Akrami:2018odb}.\\ 
	\noi Single-field slow-roll inflation is just one of several scenarios compatible with current observations. Indeed, plenty of multi-field realizations can be found  in the literature. String theory, for example, can accommodate a rich inflationary particle content including compactification moduli, axions, gauge fields, Kaluza-Klein modes \cite{Baumann:2014nda}.\\
	\noi Upon requiring that the mass of the main field driving inflation is small enough to guarantee about $60$ e-folds of expansion, any extra particle content can in principle cover a wide mass range. Very massive fields $m\gg H$ are typically integrated out, although one may look for remnants \cite{Achucarro:2012sm,Burgess:2012dz,Tolley:2009fg, Gong:2013sma, Dimastrogiovanni:2012st,Silverstein:2017zfk,Garcia-Saenz:2018ifx,Fumagalli:2019noh,Garcia-Saenz:2019njm} of such fields in late-time observables. Cosmological probes are more sensitive to lighter (i.e. more long-lived) particles, those satisfying  $m\lesssim H$, which will be the main focus of this work. It proves useful to organize any extra content according to the mass and the spin of each given particle. Primordial correlators such as the bispectrum store, in their squeezed (and generalizations thereof) configuration, key information on the mass and the spin of particles that mediate the corresponding interactions. Intriguingly, spinning fields generate a richer dynamics, including for example an extra angular dependence \cite{Noumi:2012vr,Arkani-Hamed:2015bza} (see also \cite{Jeong:2012df}) that may be searched for by current and up-coming experiments.\\
	As one considers the requirement of a consistent and \textit{predictive} theory of spinning fields in (quasi) de Sitter, the allowed mass range is dramatically reduced. Starting with $s=2$, unitarity bounds \cite{Higuchi:1986py} force massive\footnote{It is crucial to note here that the bound applies to minimally (i.e. gravitationally) coupled fields.} particles to satisfy inequalities of the schematic form $m \gtrsim H$, to the detriment of the observational prospects for signatures corresponding to spinning fields. Given that the inflationary background is not exactly de Sitter, one may hope that unitarity is less demanding on an FLRW background. This specific question has been addressed, for the case of extra spin-2 fields, in \cite{Fasiello:2013woa}: although weakened in FLRW, a consistent cosmological evolution  leads to a unitarity (or ``Higuchi'') bound of the same form. In the absence of fine-tuning, the bound is no weaker than about one order of magnitude w.r.t. the pure dS case. Intuitively, this is expected in view of the fact that one can continuously go from FLRW to dS.\\
	\noi The one implicit assumption in the above results is that the extra fields are minimally coupled, i.e. only coupled through gravity to each other and to the inflaton field. There lies the key on how to drastically weaken unitarity bounds. Indeed, the latter stem from identifying unitary representations of the de Sitter isometry group. However, the inflationary background breaks dS isometries (the very same isometries at the heart of unitarity requirements) and so will any field with sizable coupling to the inflaton. Unitarity bounds are much weakened as a result of such non-minimal couplings and a much more interesting phenomenology ensues.  This is not surprising:
	direct coupling to a light field (the inflaton, in this case) makes much more efficient the energy exchange and enables heavier modes to become effectively lighter and considerably more long-lived.\\
	
	\noi In this work we aim to explore the signatures associated to the presence of extra particle content non-minimally coupled to the inflaton field. When considering  extra fields it is important to first verify that direct coupling does not lead to instabilities, or ghostly degrees of freedom (see e.g. \cite{Himmetoglu:2009qi}). There is an extra difficulty when it comes to higher spin ($s>2$) theories in that a fully non-linear Lagrangian formulation is still missing \cite{Sorokin:2004ie,Bouatta:2004kk}. The massive $s=2$ case is special in that a ghost-free fully non-linear formulation exists \cite{deRham:2010kj,Hassan:2011zd} and has been used in the context of inflation \cite{Biagetti:2017viz,Dimastrogiovanni:2018uqy,Goon:2018fyu} (see also \cite{Fujita:2019tov} and references therein for an approach with a related model).\\
	\noi In this work we shall adopt a specific approach \cite{Bordin:2018pca} to non-minimal coupling during inflation which is the natural generalization of the effective field theory (EFT) of single field (or clock) inflation of \cite{Cheung:2007st}. This is an EFT for the fluctuations around a given FLRW solution.
	The advantage intrinsic to this formalism is that the EFT will span almost the entire space of possible signatures, yet it is possible to implement several consistency checks so that the theory is free of well-known pathologies such as e.g. gradient instabilities \cite{Bordin:2018pca}.\\ \noi We will focus in particular on the gravity sector of an inflationary theory with extra spinning fields. 	The degrees of freedom associated to spinning particles can source gravitational waves (GW) already at linear order, and the fact that these extra modes can be light enhances their (integrated) effect on late-time observables. Remarkably, the sourced contribution can be the leading one and may dramatically alter the properties of the signal w.r.t. the vacuum dominated scenario\footnote{In single-field slow-roll inflation the leading contribution to the primordial GW power spectrum stems from the homogeneous solution to the wave equation.}. Indeed, in contradistinction to the slightly red-tilted GW signal of single-field slow-roll models, multi-field set-ups enable a (strong) scale dependence in the tensor power spectrum, spanning from bump-like features to a purely blue-tilted signal. We first review  how an extra spin-2 field can source the GW spectrum at tree level \cite{Bordin:2018pca} and then show how a time-dependent speed of sound for the extra modes delivers a blue-tilted spectrum\footnote{See also e.g. \cite{Cai:2015yza,Cai:2016ldn} for more work on time-varying sound speed for the tensor modes.} at reach for the  Laser Interferometer Space Antenna (LISA). We combine bounds from  the model inner consistency checks with those originating from (i) the upper limit on the tensor-to-scalar ratio $r$ at CMB scales and scalar/mixed non-Gaussianities in the same regime (ii) ultracompact minihalos, (iii) primordial black holes, (iv) big bang nucleosynthesis, and (v) the Laser Interferometer Gravitational-Wave Observatory (LIGO). \\
	
\noi This work is organized as follows.~In Section \ref{sec:themodel} we briefly review the effective theory approach and derive the scalar and tensor spectral indices in the case of time-dependent sound speeds for the helicity modes of an extra spin-2 field. In Section~\ref{sec:bounds} we discuss the theoretical requirements on the EFT Lagrangian parameters alongside current experimental constraints. These are later employed in Section \ref{sec:analysis} to define current exclusion limits on the parameter space and to explore LISA detection and constraining power on this very general set-up.
We conclude with Section~\ref{sec:conclusions}.

\section{The effective theory approach}
\label{sec:themodel}

\begin{figure}
\centering
\begin{minipage}{.45\textwidth}
  \centering
  \includegraphics[width=1\linewidth]{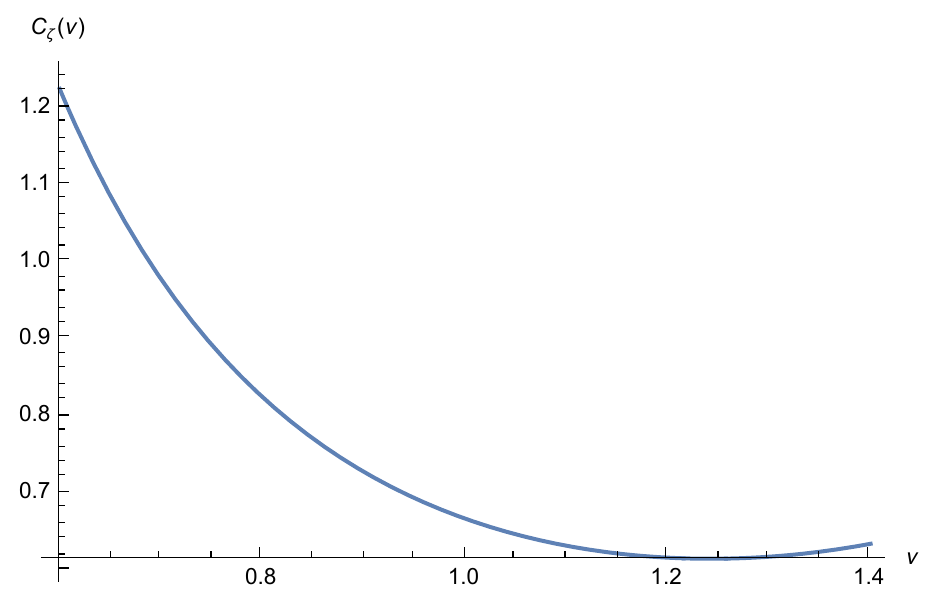}
  \caption{$C_\zeta (\nu)$.}
  \label{fig:c_scalar}
\end{minipage}
\begin{minipage}{.45\textwidth}
  \centering
  \includegraphics[width=1\linewidth]{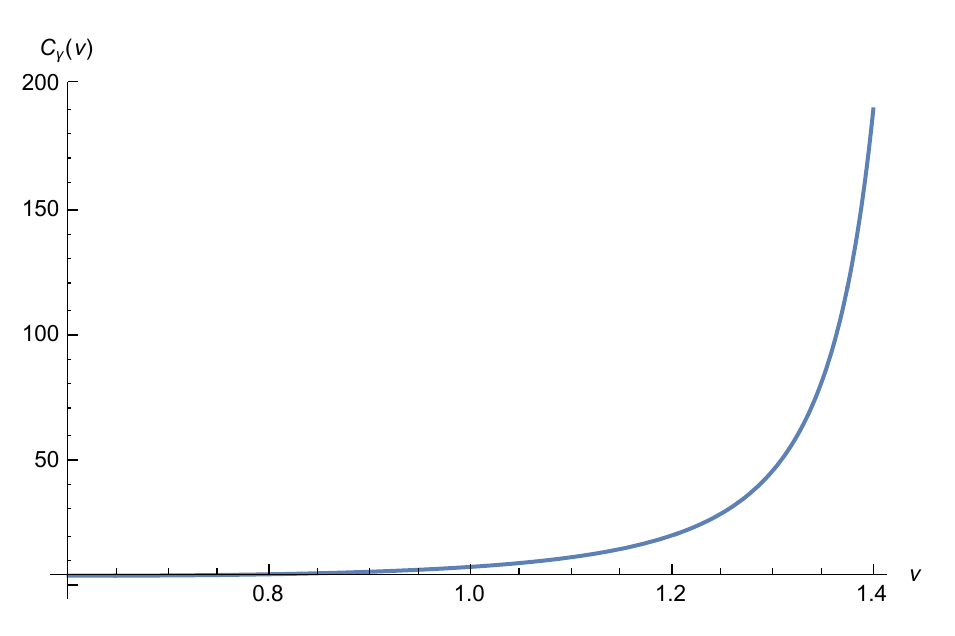}
  \caption{$C_\gamma (\nu)$.}
  \label{fig:c_tensor}
\end{minipage}
\end{figure}

The unitarity bounds that prohibit a large fraction of the mass range for fields with spin $s\geq 2$ stem from the common notion of particles as unitary irreducible representations of the spacetime isometry group. The fact that inflation does not correspond to dS, but rather to quasi-de Sitter background, points to a natural way around stringent unitarity requirements. The inflaton itself breaks dS isometries for the simple reason that inflation needs a ``clock'' for the accelerated expansion to eventually come to an end.  Demanding unitarity on extra fields in quasi dS space turns out to enforce qualitatively similar constraints to the de Sitter case as long as the extra content is only minimally, i.e. gravitationally, coupled to the inflaton field. The key step is then to \textit{directly} couple spinning particles to the inflaton. This has been implemented in several well-known (classes of) models, including axion-inflation.\\
\noi Given a specific set-up one can then work out the corresponding effects of non-minimal coupling on late-time observables. In this work we shall adopt a different perspective, namely that of \cite{Bordin:2018pca}, which is an extension of the works in \cite{Cheung:2007st,Senatore:2010wk}. The set-up of \cite{Cheung:2007st} uses the Stueckleberg trick to make manifest the goldstone boson of the (spontaneously) broken time-reparametrization invariance. For sufficiently high energy, the dynamics of the system if fully captured by the goldstone boson $\pi$, related to the curvature fluctuations $\zeta$ by $\zeta \simeq -H \pi$. It is then natural to consider extra fields in this framework \cite{Senatore:2010wk}.\\ \noi The extension to extra spinning fields is somewhat more complex. It relies on the fact that one can classify the extra field content, as is typical in the case of non-linearly realised symmetries, as representations of the unbroken group. The unbroken symmetries being rotations, it is  straightforward that particles of different spin will have a different description (as a three-vector, a three-tensor and so on). In the case of interest for us, that of an extra spin-2, the five propagating degrees of freedom are described by the traceless symmetric tensor $\Sigma^{ij}$ which is ``embedded'' as the four-tensor $\Sigma^{\mu\nu}$, whose $(0,0)$ and $(0,i)$ components are:
\bea
\Sigma^{00}= \frac{\p_i \pi \p_j \pi}{(1+\dot{\pi})^2} \Sigma^{ij}\quad ; \qquad \Sigma^{0j}= -\frac{\p_i \pi }{1+\dot{\pi}} \Sigma^{ij}\quad .
\eea
The effectively light states described by $\Sigma$ have their couplings with the inflaton prescribed by the fact that broken symmetries are non-linearly realized. An explicit example is provided in the quadratic and cubic interactions for $\sigma^{ij}= a^2 \Sigma^{ij}$ below:

\begin{equation}
\label{action}
\begin{split}
\mathcal{S} &\supseteq \mathcal{S}_{\text{free}}^{(2)}+\mathcal{S}_{\text{int}}^{(2)} +\mathcal{S}_{\text{int}}^{(3)} \\
&=\frac{1}{4}\int dt\,  d^3x\, a^3 \Big[(\dot{\sigma}^{ij})^2 -c_2^2a^{-2}(\partial_i \sigma^{jk})^2 -\frac{3}{2}(c_0^2-c_2^2)a^{-2}(\partial_i \sigma^{ij})^2 -m^2(\sigma^{ij})^2 \Big] \\
&\;\; +\, \int dt \, d^3x\, a^3 \Big[ -\frac{\rho}{\sqrt{2 \epsilon_1} H} a^{-2} \partial_i \partial_j \pi_c \sigma^{ij} +\frac{1}{2} \rho \dot{\gamma_c\,}_{ij} \sigma^{ij}  \Big]  \; \\
&\;\;-\, \int dt \, d^3x\, a^3 \Big[ \frac{\rho}{2\epsilon_1 H^2 M_{\rm Pl}}a^{-2} (\p_i\pi_c \p_j\pi_c \dot{\sigma}^{ij}+2H \p_i\pi_c \p_j\pi_c {\sigma}^{ij} ) +\mu(\sigma^{ij})^3 +\dots \Big]
\end{split}
\end{equation}
where $\epsilon_1\equiv-\dot{H}/H^2$, $m$ is the mass of the spin-$2$ field and $\pi_c\equiv\sqrt{2 \epsilon_1} H M_{\rm Pl}\, \pi$ is the canonically normalized goldstone boson, linearly related to $\zeta$. Note that non-linear diffs also dictate couplings with tensor fluctuations $\gamma_{ij}$, canonically normalized as ${\gamma^c\,}_{ij}=M_{\rm Pl} \gamma_{ij}$. The quantity $\rho$ is a coupling constant with mass dimension one, and the sound speeds for the 0, 1 and 2-helicity components of $\sigma$ are indicated as $c_{0}$, $c_{1}$ and $c_{2}$. These satisfy the following relation \cite{Bordin:2018pca}:
\begin{equation}
\label{sound_speeds_relation}
c_1^2=\frac{1}{4}c_2^2+\frac{3}{4}c_0^2 \;.
\end{equation}
The interactions in $\mathcal{S}_{\text{int}}^{(2)}$ generate tree-level contributions to the scalar and tensor power spectra, whose amplitudes are sensitive to the sound speeds $(c_{0},\,c_{2})$, and to the magnitude of the $\rho/H$ coupling. In this work, we shall allow for a time-dependence in the sound speeds and explore its implications at the level of the power spectra.  
The scale-dependence for the scalar and tensor power spectra is given by:
\begin{subequations}
	\label{powerlawscale}
	\begin{gather}
		\label{powerlawscale:scalar}
		\mathcal{P}_\zeta (k) =\frac{H^2}{8 \pi^2 M_{\rm Pl}^2 \epsilon_1}\left( \frac{k}{k_\star}\right)^{n_s^{(v)}-1}+ \frac{H^2}{8 \pi^2 M_{\rm Pl}^2 \epsilon_1^2} \frac{C_\zeta (\nu)}{{c_0}^{2\nu}}  \left(\frac{\rho}{H} \right)^2 \left( \frac{k}{k_\star}\right)^{n_s^{(\sigma)}-1} \;,\\
		\label{powerlawscale:tensor}
		\mathcal{P}_\gamma (k)=\frac{2 H^2}{\pi^2 M_{\rm Pl}^2 } \left( \frac{k}{k_\star}\right)^{n_t^{(v)}}+ \frac{2 H^2}{\pi^2 M_{\rm Pl}^2 } \frac{C_\gamma (\nu)}{ {c_2}^{2\nu}} \left(\frac{\rho}{H} \right)^2 \left( \frac{k}{k_\star}\right)^{n_t^{(\sigma)}}   \;,
	\end{gather}
\end{subequations}
where $k_\star$ is a pivot scale set at $0.05\;\mbox{Mpc}^{-1}$. The first contribution on the right-hand side of Eq.~(\ref{powerlawscale:scalar}) is due to vacuum fluctuations, the second one is sourced by $\sigma$ (similary for Eq.~(\ref{powerlawscale:tensor})). The scalar and tensor spectral indices are given by
\begin{subequations}
	\begin{gather}
		\label{tilt vacuum}
		n_s^{(v)}-1= -2\epsilon_1-\epsilon_2    \;, \\
		n_s^{(\sigma)}-1= -2\epsilon_2 -2\nu s_0 -\frac{m^2}{H^2}\epsilon_1 \frac{1}{\nu}\Big( \frac{\partial}{\partial \nu} \ln C_\zeta (\nu)-2\ln c_0\Big)\;,  \\
		n_t^{(v)}=-2\epsilon_1\;,  \\
		n_t^{(\sigma)}=-2\nu s_2 -\frac{m^2}{H^2}\epsilon_1 \frac{1}{\nu}\Big( \frac{\partial}{\partial \nu} \ln C_\gamma (\nu)-2\ln c_2\Big) \;.
	\end{gather}
\end{subequations}
In the above,  $\nu=\sqrt{9/4-(m/H)^2}$, $\epsilon_{i+1}\equiv \dot{\epsilon_i}/(H\epsilon_i)$ and the time-dependence of the sound speeds is described in terms of the parameters $s_n\equiv \dot{c_n}/(H c_n)$, with $n=0,1,2$. We take the parameter $\rho$ to be a constant. The functions $C_\zeta (\nu)$ and $C_\gamma (\nu)$ can be computed analytically for $c_0\ll1$ and $c_2\ll1$ \cite{Bordin:2018pca} and are represented in Figs.~(\ref{fig:c_scalar}) and (\ref{fig:c_tensor}) in the mass range $\nu \in [3/5,7/5]$.

\section{Current and future bounds}
\label{sec:bounds}
Let us elaborate on the bounds on the scalar and tensor power spectra that are employed in Section~\ref{sec:analysis} to constrain the parameter space of the effective Lagrangian. We refer the interested reader to a similar analysis carried out in \cite{Bartolo:2016ami}. There, the starting point is the specific model in \cite{Fujita:2014oba} (see also \cite{Biagetti:2013kwa,Biagetti:2014asa}), with an additional feature: the time-dependence of the speed of sound for the scalar spectator field has been switched on. In what follows, theoretical requirements and experimental constraints  will  be combined to draw the current exclusion regions in the $(\rho/H,\,|s_{2}|)$ plane of the EFT Lagrangian parameter space and to identify LISA's constraining  power.

\subsection{Consistency requirements}
Besides being generally defined in the $(0,1]$ interval, the sound speeds in the EFT framework are subject to additional theoretical (and observational) constraints. One such bound, $c_{n}\gtrsim 10^{-3}$, arises from perturbativity requirements, in particular from requiring that the one loop $\sigma$-sourced corrections are smaller than the tree-level power spectra \cite{Bordin:2018pca}.\\ General consistency of the perturbative treatment also traslates into bounds on the coupling constants, $\rho/H \ll \sqrt{\epsilon_1}$ and $\rho/H \ll 1$ (weak-mixing regime). These ensure that $\mathcal{S}_{\text{int}}^{(2)}$ can be treated as a (small) correction to the kinetic Lagrangian. \\
An additional constraint on $\rho/H$ is imposed in order to avoid gradient instabilities:
\begin{equation}
\label{grad instab}
    \frac{\rho}{H} \ll \sqrt{\epsilon_1 c_0^2} \;.
\end{equation}
Eq.~(\ref{grad instab}) automatically ensures that also the previous two conditions on $\rho/H$ are satisfied, given that $c_0\leq1$ and $\epsilon_1\ll1$.  It is also important to point out that the expressions of the power spectra reported in Eqs.~\eqref{powerlawscale} are only accurate \cite{Bordin:2018pca} in the regime 
\begin{subequations}
\label{two bounds perturbativity}
\begin{gather}
\label{two bounds perturbativity:1}
\rho \ll m \;,\\
\label{two bounds perturbativity:2}
\Big(\frac{m}{H} \Big)^2 \gg \frac{1}{N(k)} \;,
\end{gather}
\end{subequations}
where $N(k)$ is the number of e-folds between the horizon crossing time of the mode with wavenumber $k$ and the end of inflation. The conditions in (\ref{two bounds perturbativity}) ensure that the perturbative result for the tensor power spectrum coincides with the one derived from a non-perturbative treatment of the $(\gamma-\sigma)$ mixing.

\subsection{Observational bounds on the scalar sector}

\noindent The two and three-point statistics of scalar perturbations are constrained on large scales by measurement of CMB anisotropies.  Constraints on scalar non-Gaussianity on these scales give rise to a lower bound on the sound speeds of the order $c_{n}\gtrsim 10^{-2}$ \cite{Bordin:2018pca}. As to the two-point statistics, Planck data \cite{Akrami:2018odb} yield a power spectrum amplitude $A_s=2.101 \times 10^{-9}$, and a tilt $n_s=0.9649$, where
\begin{equation}
\label{Planck param}
\mathcal{P}_\zeta^{\text{Planck}} (k)=A_s \left(\frac{k}{k_\star} \right)^{n_s-1} \;.
\end{equation}
In the following we will assume, for simplicity, that on large scales the scalar power spectrum is dominated by the vacuum contribution; according to Eq.~\eqref{powerlawscale:scalar}  this implies
\begin{equation}
\label{vacCMB}
\frac{\rho}{H} \ll \sqrt{\frac{\epsilon_1 c_0^{2\nu}}{C_{\zeta}(\nu)}} \Big|_{k=k_\star}\;. 
\end{equation}
As it will become clearer in the next section, within the spin-2 field mass range considered in this work Eq.~\eqref{vacCMB} and the gradient instability condition  \eqref{grad instab} are nearly equivalent. When Eq.~(\ref{vacCMB}) holds, $\epsilon_1$ can be deduced from the value of $A_s$  upon fixing the Hubble rate. One can further verify that, given the above conditions, Planck constraints on the tilt are easily satisfied. To this aim, we compute the EFT Lagrangian prediction for the parameters in (\ref{Planck param}):
\begin{equation}
\mathcal{P}_\zeta(k)=\frac{H^2}{8 \pi^2 M_{\rm Pl}^2 \epsilon_1} \left[ 1+ \frac{C_\zeta (\nu)}{{c_0}^{2\nu} \epsilon_1}  \left(\frac{\rho}{H} \right)^2 \right]     \left(\frac{k}{k_\star}\right)^{n_{s}^{\text{tot}}} \;, 
\end{equation}    
where $n_{s}^{\text{tot}}= -2\epsilon_1-\epsilon_2   +\alpha  |_{k=k_\star}$, with 
\begin{equation}
\alpha= \frac{\frac{C_\zeta (\nu)}{{c_0}^{2\nu} \epsilon_1}  \Big(\frac{\rho}{H} \Big)^2}{1+\frac{C_\zeta (\nu)}{{c_0}^{2\nu} \epsilon_1}  \left(\frac{\rho}{H} \right)^2}\left[-\epsilon_2 +2\epsilon_1-2\nu s_0 -\frac{m^2}{H^2}\epsilon_1 \frac{1}{\nu}\left( \frac{\partial}{\partial \nu} \ln C_\zeta (\nu)-2\ln c_0\right) \right]  \;.
\end{equation}
Using the observed value of the tilt to fix $\epsilon_2=-2\epsilon_1-(n_s-1)$, one concludes that the condition $\alpha\ll -2\epsilon_1-\epsilon_2$ puts a bound on the parameter space which is automatically satisfied if Eq.~(\ref{vacCMB}) holds, and provided we span the range of masses and sound speeds used throughout our analysis.

\noi The amplitude of the scalar fluctuations is constrained on small scales by CMB spectral distortions, primordial black holes (PBH) and ultra-compact mini-halos (UCMH).  Spectral distortions can be generated by dissipation of primordial perturbations through photon diffusion and are relevant in the $1-10^{4}\,\text{Mpc}^{-1}$ range \cite{Chluba:2012gq}. PBH may have formed from the collapse of (large) density perturbations and therefore constraints on their abundance result into bounds on the primordial scalar power spectrum \cite{Carr:2009jm,Josan:2009qn,Carr:2017jsz}. These are several orders of magnitude weaker compared with CMB bounds, however they are important in that they span significantly more orders of magnitude in scales (see \cite{Kalaja:2019uju} for an updated analysis). We verified that both spectral distortions and PBH constraints produce less stringent bounds on the parameter space of our EFT Lagrangian than those obtained by implementing the  gradient instability condition. These will therefore not be included in the final plots.\\
\noi UCMH are dense dark matter structures that can form from large density perturbations right after matter-radiation equality \cite{Ricotti:2009bs}. While many of the constraints on their abundance depend on assumptions regarding the nature of the dark matter particles, more general constraints can be obtained by accounting for gravitational effects, in particular lensing time-delay in pulsar timing. For this reason we apply UCMH constraints at a scale $k_{\text{UCMH}}=3\times 10^{5}\,\mbox{Mpc}^{-1}$. The value of the bound depends on whether one assumes a constant or a scale and redshift-dependent $\delta_{\text{min}}$ ($\delta_{\text{min}}$ being the minimal value of the density contrast that is required to form the UCMH structure). In the scale-independent case the bound is given by $P_\zeta(k_{UCMH})<3\times 10^{-8}$. If instead a scale-dependence is allowed the corresponding bound   \cite{Emami:2017fiy} is $P_\zeta(k_{\text{UCMH}})<10^{-6}$. In Section~\ref{sec:analysis} we implement both and refer to them as ``UCMH ($\delta$ const)'' and ``UCMH ($\delta$ scale dep)''. 

\subsection{Observational bounds on the tensor sector}

The current upper limit on the tensor-to-scalar ratio is given by $r<0.056$, at a pivot scale $ k_r=0.002 \, \mbox{Mpc}^{-1}$ \cite{Akrami:2018odb}. For the set-up under scrutiny, this bound gives
\begin{equation}
\label{rbound}
\begin{split}
 \mathcal{P}_\gamma^{(v)}(k_r)+\mathcal{P}_\gamma^{(\sigma)}(k_r)< 1.3\times 10^{-10} \;,
\end{split}
\end{equation}
where the superscripts $(v)$ and $(\sigma)$ indicate respectively the vacuum and sourced contributions in Eq.~\eqref{powerlawscale:tensor}. For each configuration analyzed in Sec.~\ref{sec:analysis}, characterized by fixed values of $\{H,\,m\}$ and initial conditions for the sound speeds, Eq.~\eqref{rbound} will generate an exclusion line in the $(\rho/H,\,|s_2|)$ plane. In the case of a tensor power spectrum  dominated by vacuum fluctuations on large scales, the constraint on $r$ corresponds to a maximum value $H^{\text{max}}=6.13\times 10^{13}\,\mbox{GeV}$ for the Hubble rate during inflation. In Section~\ref{sec:analysis} we will derive exclusion limits on the parameter space for the following values: $H=\{10^{12}\,\mbox{GeV},\,10^{13}\,\mbox{GeV}, \,6.1\times10^{13}\,\mbox{GeV}\}$.\\ 

\noindent In addition to constraints on the tensor power spectrum, measurements of CMB anisotropies reflect on tensor non-Gaussianity in the form of a lower bound on the helicity-2 sound speed, $c_{2}\gtrsim 10^{-2}$ \cite{Dimastrogiovanni:2018gkl}. This bound will restrict the range of possible values for the initial (i.e. large-scale) $c_{2}$.\\

\noindent Inflationary tensor fluctuations contribute to the present gravitational waves energy density, $\Omega_{\text{GW}}$, as follows
\begin{equation}
\label{omega}
\Omega_{\text{GW}}(k,\tau_0)=\frac{1}{12}\,\left(\frac{k}{a_0 H_0} \right)^{2}P_{\gamma}(k) \, T^2(k,\tau_0) \;.
\end{equation}
Here quantities with the label "$0$" are evaluated at present time, $\tau$ is the conformal time, $H_0=100h \, \mbox{km}/\mbox{s}/\mbox{Mpc}$ is the Hubble rate today (in the following we use $h=0.674$ \cite{Aghanim:2018eyx}), and $T(k, \tau_0)$ the transfer function. We consider a standard reheating scenario where inflation is initially followed by a matter-dominated phase and, once reheating is completed, the universe becomes radiation-dominated. In this context the transfer function is given by (see e.g. \cite{Kuroyanagi:2014nba, Bartolo:2016ami})
\begin{equation}
\label{transfer}
T(k,\tau_0)=\frac{3\, \Omega_m \,j_1(k \tau_0)}{k \tau_0} \sqrt{1.0+1.36\Big(\frac{k}{k_{\text{eq}}} \Big)+2.50 \Big(\frac{k}{k_{\text{eq}}} \Big)^2} \;,
\end{equation} 
where $\Omega_m=0.315$ \cite{Aghanim:2018eyx} is the matter energy-density today, $j_1(k\tau_0)$ is the first spherical Bessel function and $k_{\text{eq}}=7.1\times 10^{-2} \,\Omega_m \,h^2 \,\mbox{Mpc}^{-1}$ is the horizon scale at radiation-matter equality. We assume the standard $\Lambda \mbox{CDM}$ cosmological model, with $\Omega_K$ and $\Omega_r$ negligible, so that
\begin{equation}
\tau_0=\int_0^\infty\frac{dz}{a_0\, H_0 \, \sqrt{\Omega_\Lambda +\Omega_m (1+z)^3}} \;.
\end{equation}
\noi Existing bounds on $\Omega_{\text{GW}}(k)$, besides those from CMB anisotropies, are provided by pulsar timing arrays (PTA), advanced LIGO, big-bang nucleosynthesis, and by the CMB monopole. \\
Data from the second observing run of advanced LIGO, combined with the results of the first run, can be used to place upper limits on $\Omega_{\text{GW}}$ for a background which is frequency-independent in the LIGO frequency band. The bound is given by \cite{LIGOScientific:2019vic}
\begin{equation}
\label{LIGObound}
\Omega_{\text{GW}}<6.0 \times 10^{-8} 
\end{equation} 

\noi As for all bounds listed in this section, we translate this limit into a constraint in the $(\rho/H,\,|s_2|)$ space. In particular, for each configuration tested in Sec.~\ref{sec:analysis}, we replace the expression of the tensor power spectrum \eqref{powerlawscale:tensor} into Eq.~\eqref{omega}, and derive the LIGO exclusion line from (\ref{LIGObound})\footnote{In principle, we should use a bound from a specific search for the spin$-2$ signal in LIGO data. However, as shown in Sec.~\ref{sec:analysis}, the LIGO bound is never the strongest one in the parameter space, hence we find it safe to use the constraint given on a flat signal.}.\\
\noi Measurements of the abundance of the primordial light elements constrain the number of effective massless degrees of freedom at the onset of nucleosynthesis. This bound is weaker than the LIGO constraint \eqref{LIGObound}. It can be  verified that for a monotonically growing primordial power spectrum as $\mathcal{P}_{\gamma}^{(\sigma)}$, the BBN exclusion line always sits above the LIGO line in the $(\rho/H,\,|s_2|)$ plane. The same conclusion applies to the CMB monopole and PTA bounds, which are therefore not represented in the plots of Sec.~\ref{sec:analysis}.
\begin{figure}[t]
	\centering
	\includegraphics[scale=0.45]{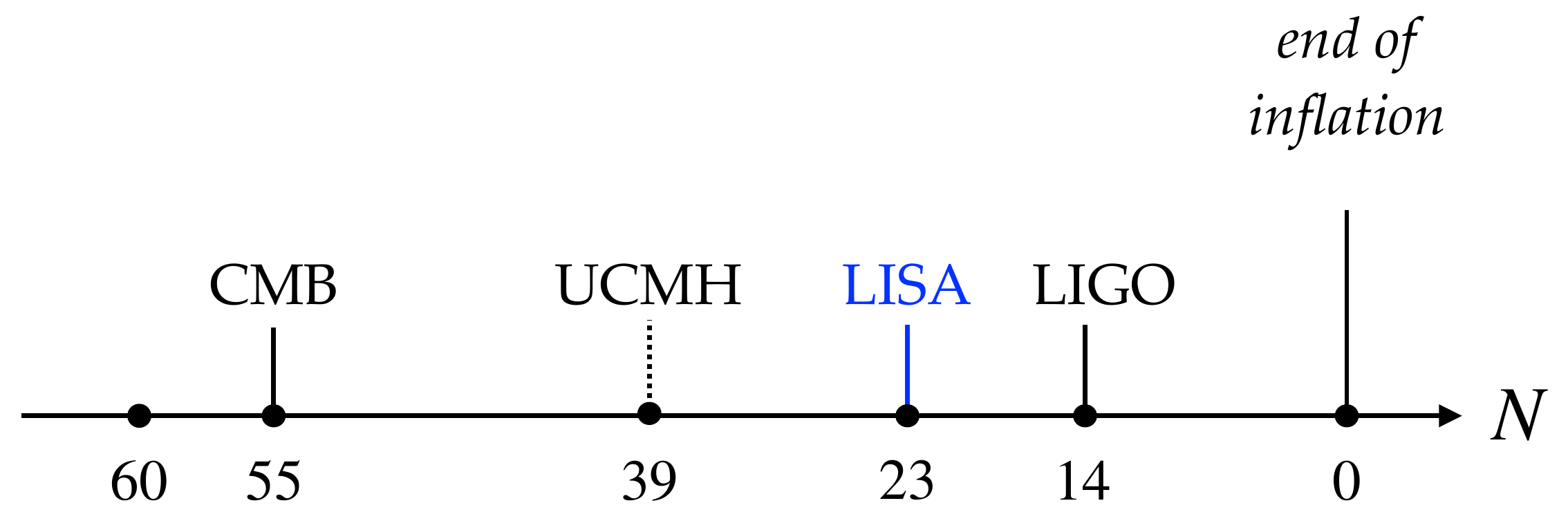}
	\caption{Sketch of $N_{tot}\simeq60$ e-folds of inflationary expansion ($H=10^{12}\, \mbox{Gev}$), where we highlight relevant stages corresponding to experimental bounds discussed in Sec.~\ref{sec:bounds}. The LISA line is highlighted in blue.}
	\label{fig:Nefolds}
\end{figure}

\noi Besides existing observational bounds on $\Omega_{\text{GW}}$, we also consider LISA expected sensitivity limits. 
The  duration of the mission will be 4 years, with a possible 6-year extension, and LISA arms will be $2.5\times 10^6 \,{\rm Km}$ long. The data taking efficiency of the mission is expected to be $\sim 75\%$ of the nominal time, because of operations needed for the antenna mantainance. As a result, the 4-year mission will effectively produce 3 years of data. The most updated LISA strain sensitivity curve can be found in \cite{Caprini:2019pxz}, where the power law sensitivity curve is derived following \cite{Thrane:2013oya}, for a signal-to-noise ratio $\rm{SNR}=10$. In the following, we pick as reference scale for LISA the one that minimizes the sensitivity curve calculated in \cite{Caprini:2019pxz}, $k_{\text{LISA}}\simeq 1.79 \times 10^{12}\, \mbox{Mpc}^{-1}$, which corresponds to a sensitivity value $\Omega_{\text{LISA}}\simeq 4.12\times 10^{-13}$. In order to be detectable by LISA, the energy-density associated with a gravitational wave mode must overcome LISA's sensitivity curve at the same scale, $\Omega_{\text{GW}}(k)>\Omega_{\text{LISA}}(k)$. This condition is used to generate the LISA line in $(\rho/H,\,|s_2|)$ plane (see plots in Section~\ref{sec:analysis}).\\

\noi We provide in Fig.~\ref{fig:Nefolds} a pictorial representation of $N_{tot}\simeq 60$ e-folds of inflationary expansion for $H=10^{12}\, \mbox{GeV}$ and highlight the stages at which the relevant scales left the horizon.  

\section{Examples of time-dependent sound speeds}
\label{sec:analysis}
The parameters $s_n\equiv \dot{c_n}/(H c_n)$, where $n=0,1,2$  for the helicity 0, 1 and 2 of $\sigma$, quantify the time-dependence of the sound speeds. From Eq.~\eqref{sound_speeds_relation} one finds 
\begin{equation}
\label{relation_sss}
s_0= \frac{4}{3}\frac{c_1^2}{c_0^2}s_1-\frac{1}{3}\frac{c_2^2}{c_0^2}s_2 \;.
\end{equation}

\noindent In Table~\ref{tab:table_solutions}, we report some of the solutions to \eqref{relation_sss}. A negative (positive) $s_{n}$ produces a decreasing (increasing) sound speed for the corresponding helicity, hence a power spectrum that grows (decreases) towards small scales. For tensor perturbations, this implies the existence of a gravitational wave signal potentially detectable with interferometers. The  solutions with $s_{2}<0$, which we shall refer to as (1.a), (2.a), and (3.a), are highlighted in yellow in Table~\ref{tab:table_solutions}.

\noi Naturally, we anticipate that the region of parameter space corresponding instead to an enhanced scalar power spectrum at small scales, $s_{0}<0$, will be more constrained, given also the absence of  gradient instabilities enforced by Eq.~\eqref{grad instab}.  \\

\begin{table}
	\centering
	\begin{tabular}{| c | c | c | c | c |} 
		\hline
		$s_0$ & $s_1$ & $s_2$&      &   case \\ [0.5ex] 
		\hline \hline
		\multirow{2}{*}{$s_0=-\frac{1}{3}\frac{c_2^2}{c_0^2}s_2$} &\multirow{2}{*}{$s_1=0$}& \multirow{2}{*}{$s_2$} & $s_2<0, s_0>0$ & \cellcolor{yellow} $(1.\mbox{a})$ \\ \cline{4-5}
		&&& $s_2>0, s_0<0$ & $(1.\mbox{b})$ \\
		\hline
		\multirow{2}{*}{$s_0=0$}&\multirow{2}{*}{$s_1=\frac{1}{4}\frac{c_2^2}{c_1^2}s_2$} &  \multirow{2}{*}{$s_2$} & $s_2<0, s_1<0$ & \cellcolor{yellow} $(2.\mbox{a})$\\ 
		\cline{4-5}
		&&& $s_2>0, s_1>0$& $(2.\mbox{b})$ \\
		\hline
		\multirow{2}{*}{$s$} &\multirow{2}{*}{$s$} & \multirow{2}{*}{$s$} & $s<0$ &\cellcolor{yellow} $(3.\mbox{a})$ \\
		\cline{4-5}
		&&&$s>0$& $(3.\mbox{b})$ \\
		\hline
	\end{tabular}
	\caption{\label{tab:table_solutions} Table displaying some of the phenomenologically interesting solutions of Eq.~\eqref{relation_sss}. The fourth column indicates the signs of the solutions. The cases highlighted in yellow are analysed in Secs.~\ref{sec:case1a}, \ref{sec:case2a} and \ref{sec:case3a}. }
\end{table}

\subsection{Case (1.a): constant $c_1$}
\label{sec:case1a}
Let us consider the solution  $\{s_0=-\frac{1}{3}\frac{c_2^2}{c_0^2}s_2>0, \, s_1=0,\,s_2<0\}$. In addition, let us assume $s_{2}=\text{constant}$ for simplicity. The time evolution of $c_{2}$ reads
\begin{equation}
\label{c_2time}
c_2(t)=c_{2 i}\, e^{s_2 N(t,\,t_{i})} \;,
\end{equation}
where $N(t,\,t_{i})\equiv \int_{t_{i}}^{t} H (t') \, dt'$ is the number of e-folds between a given reference time $t_{i}$ (we take this to be the time at which our current  observable universe exited the horizon) and $t$, with $c_{2 i}\equiv c_{2}(t_{i})$. We consider benchmark values $c_{2i}=\{10^{-2},10^{-1},1\}$. This set of initial conditions lies comfortably within the range allowed by perturbativity requirements ($c_{2}\gtrsim 10^{-3}$) and CMB constraints on non-Gaussianity ($c_{2}\gtrsim 10^{-2}$). Enforcing perturbativity throughout the scales results in a lower limit for $s_2$. Given that our $k$ regime of interest is that where the expressions in Eq.(\ref{powerlawscale}) are a good approximation for the power spectra, this identifies, via Eq.~\eqref{two bounds perturbativity:2}, an upper value for $k$ that we indicate as $k_F$.  We obtain $|s_{2}|^{\text{max}}$ by requiring that $c_{2}(k_{F})$ saturates the perturbativity bound.

\noindent In the configuration (1.a), $c_1$ is constant and $c_0$ increases as inflation proceeds. Using Eq.~\eqref{sound_speeds_relation} one obtains the time-evolution of $c_0$ 
\begin{equation}
\label{c0(k)}
c_0(k)=\sqrt{\frac{4}{3}c_1^2 -\frac{1}{3}c_2(k)^2} \;.
\end{equation}

\begin{figure}[H]
	\centering
	\includegraphics[scale=0.67]{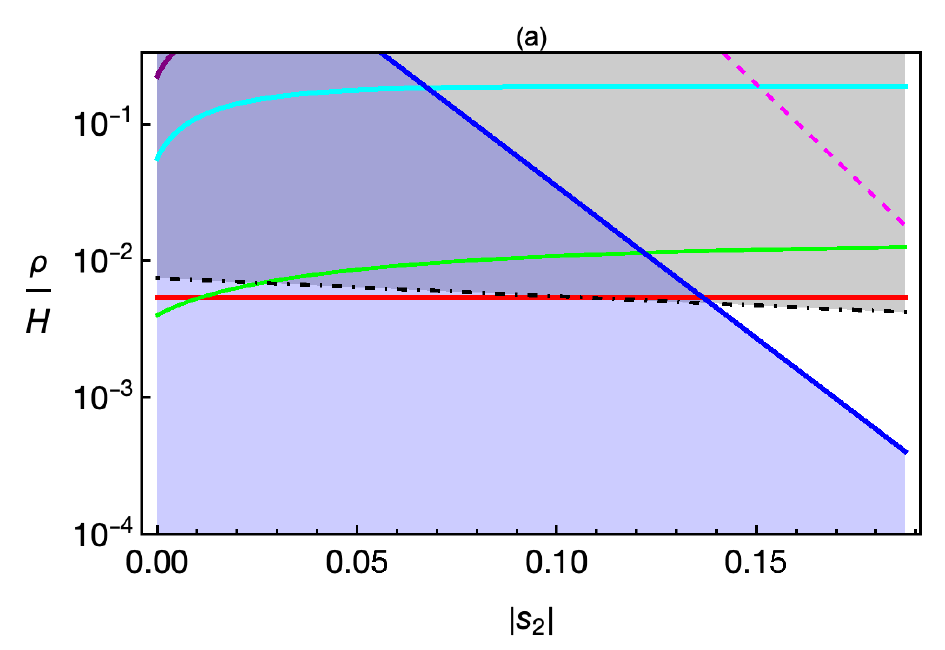} \quad \includegraphics[scale=0.67]{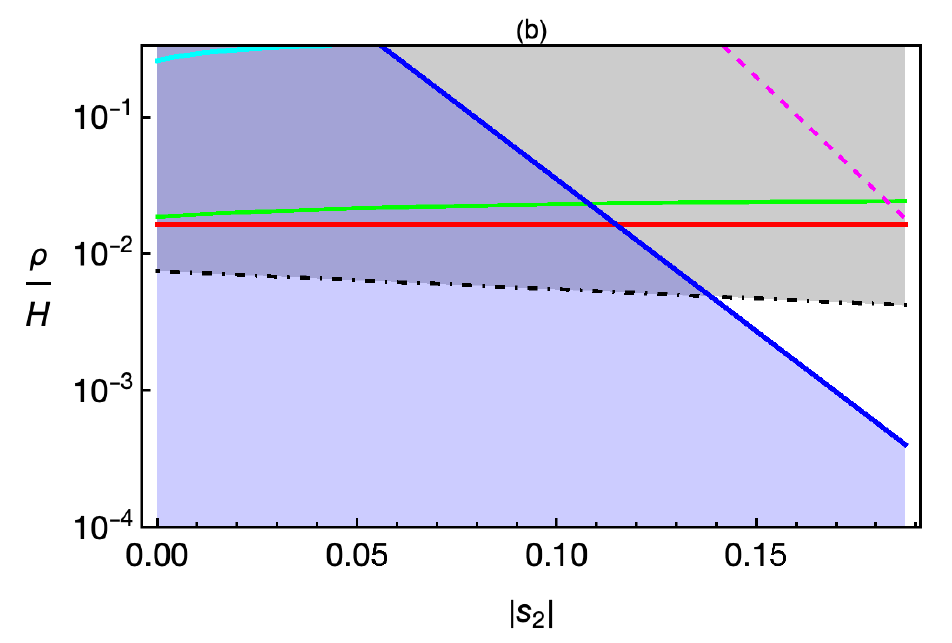} \\ [\baselineskip]
	\includegraphics[scale=0.76]{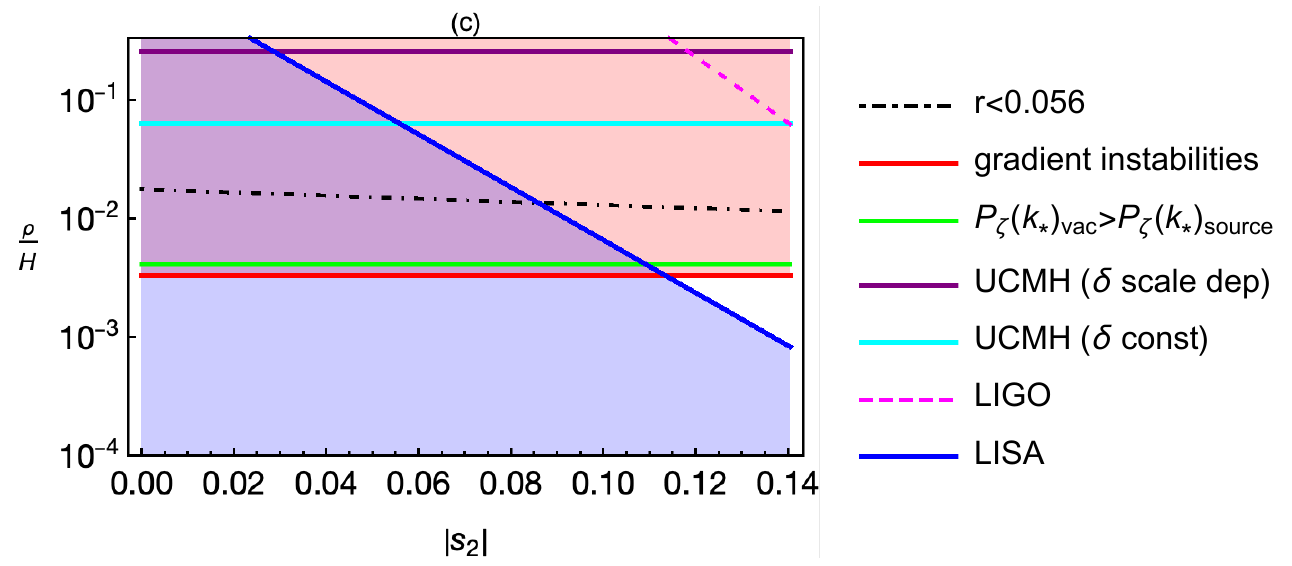} 
	\caption{ Case (1.a). We consider the $\left(\rho/H, |s_2|\right)$ plane of the effective theory parameter space for some of the configurations that can be probed with LISA. We identify the strongest bounds among those considered and shade with the corresponding color the area of parameter space that is (or would be, in the case of detection by LISA) excluded.   Top left (a): the configuration $\{H=6.1 \times 10^{13}\, \mbox{GeV},\, m/H=0.54, \, c_1=0.55, \, c_{2i}=1\}$ is displayed. Top right (b): $\{H=6.1 \times 10^{13}\, \mbox{GeV},\, m/H=0.54, \, c_1=0.85, \, c_{2i}=1\}$. Bottom (c): $\{H=10^{13}\, \mbox{GeV},\, m/H=0.54, \, c_1=0.85, \, c_{2i}=0.1\}$. Bounds discussed in Sec.~\ref{sec:bounds} that are weaker than $\rho/H<1/3$ are not captured by Fig.~\ref{fig:case1a} or the following plots, they have nevertheless been taken into account in our analysis.}
	\label{fig:case1a}
\end{figure}

\noi Requiring $c_{0}$ to be a real quantity, alongside the perturbativity and subluminality conditions on the sound speeds throughout their evolution, defines, for each $c_{2i}$ value, the corresponding range for $c_{1}$. One can easily verify that values of $c_1$ in the interval $[0.55, 0.85]$ are allowed for all chosen $c_{2i}$ values. We proceed by selecting a number of sample values for the set in $\{H, \, m/H, \, c_1, \, c_{2i}\}$ and applying the constraints described in Section~\ref{sec:bounds} to obtain (current and future) exclusion lines in the $(\rho/H,\,|s_2|)$ plane. Let us point out that, rather than describing the time-evolution of $c_0(k)$ using the leading-order term in its Taylor expansion (as would be the case for the expression in \eqref{powerlawscale:scalar}), we derive the exact scale dependence of the power spectrum by using Eq.~\eqref{c0(k)} directly.\\
We consider the masses $m/H=\{1.37,\,1.12,\,0.54\}$, identify in each plot the strongest among all the bounds and shade with its corresponding color the area of parameter space that is excluded. We also shade in blue the area which would be excluded by LISA in case of detection.\\ 
We show in Fig.~\ref{fig:case1a} the parameter space associated to configurations that are within reach for LISA. Among the mass values of the spin-2 particle which we test,  the lowest, $m/H=0.54$,  allows for a detectable signal. This reflects the fact that, the lighter the spin-2, the stronger its effect on the tensor power spectrum at small scales.

\subsection{Case (2.a): constant $c_0$}
\label{sec:case2a}
\begin{figure} [t]
	\centering
\includegraphics[width=0.47\linewidth]{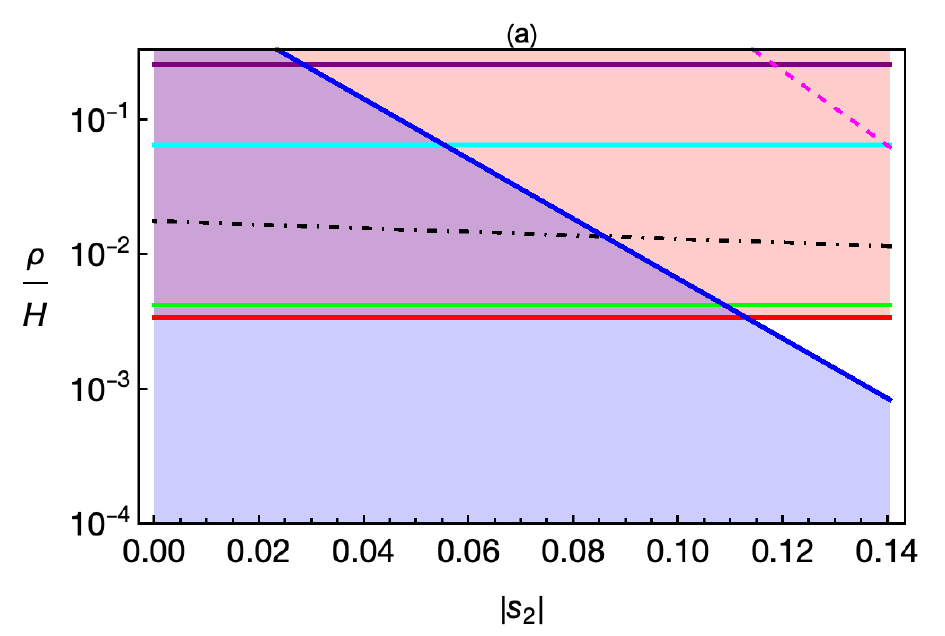} \\ \includegraphics[width=0.47\linewidth]{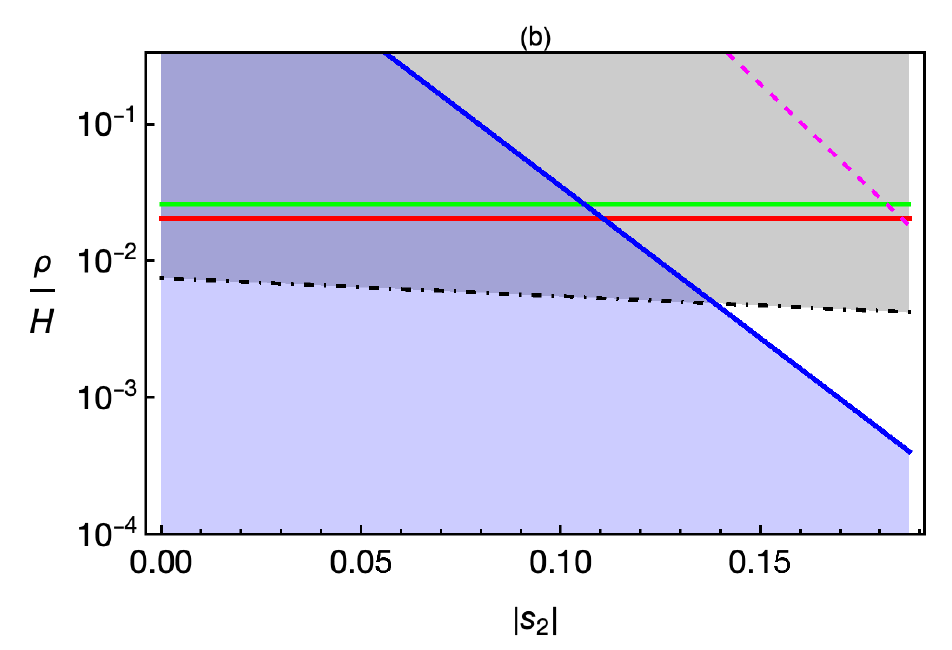}  
	\caption{ Case (2.a). The  $(\rho/H, |s_2|)$ plane for configurations at reach with LISA. In the top panel (a), the configuration $\{H=10^{13}\, \mbox{GeV},\, m/H=0.54, \, c_0=1, \, c_{2i}=0.1\}$ is displayed; in the bottom panel (b) we show the plot  corresponding to $\{H=6.1\times 10^{13}\, \mbox{GeV},\, m/H=0.54, \, c_0=1, \, c_{2i}=1\}$. Conventions for colors and line codes are as in Fig.~\ref{fig:case1a}.}
	\label{fig:case2a}
\end{figure}

Let us now consider the solution $\{ s_0=0,\, s_1=\frac{1}{4}\frac{c_2^2}{c_1^2}s_2<0,\, s_2=\text{constant}<0\}$. The time-evolution of $c_2$ is similar to case (1.a), $c_0$ is constant and $c_1$ decreases in time as
\begin{equation}
\label{c_1 evolution}
c_1(k)=\sqrt{\frac{3}{4}c_0^2+\frac{1}{4}c_2(k)^2} \;.
\end{equation}
One can verify that values $c_0=\{10^{-2},10^{-1},1\}$ and $c_{2i}=\{10^{-2},10^{-1},1\}$ guarantee subluminal propagation also for the helicity-1 mode, in addition to preserving perturbativity bounds. We consider a number of sample configurations $\{H, m/H,\, c_0,\, c_{2i}  \}$ and represent in Fig.~\ref{fig:case2a} the constraints on the parameter space for those that are potentially observable with LISA.\\

\subsection{Case (3.a): monotonically decreasing sound speeds }
\label{sec:case3a}
We explore here the solution $\{s_0=s_1=s_2=s<0\}$: in this case all the sound speeds decrease over time, and we capture their dynamics by means of Eq.~\eqref{c_2time}. For simplicity, we focus on the initial conditions $c_{0i}=c_{1i}=c_{2i}=1$. In this case also the scalar power spectrum is blue-tilted. It is straightforward to conclude that in this case none of the configurations $\{H,\, m/H\}$ tested corresponds to a signal above the sensitivity limits of LISA. Indeed, what is behind the most stringent constrain in this case is the fact that $c_0$ decreases with time.  As a consequence, the line representing the gradient instabilities bound of Eq.\eqref{grad instab} bends downwards (i.e. the bound gets stronger) in the $(\rho/H, |s_2|)$ plane  as $|s_2|=|s_0|$ increases, preventing any crossing with the LISA curve.

\subsection{Additional remarks}
Before drawing our conclusions, we remind the reader that an upper bound is imposed on $|s_2|$ in each configuration from theoretical consistency. This limits the region accessible to interferometers: the lines representing LIGO and LISA bounds bend downwards as $|s_2|$ increases.\\

\noi It is also worth remarking that the parameter space of some of the configurations analyzed for cases (1.a) and (2.a) ends up being rather similar (see Table \ref{tab:similar_cases}). This should not come as a surprise in light of Eq.~(\ref{sound_speeds_relation}). We also stress that we have focused our analysis mainly (i) on two specific observables, namely scalar and tensor power spectra, and (ii) considered a bi-dimensional sub-region of the entire parameter space, the $(\rho/H, |s_2|)$ plane. Extending the dimensionality of the parameter space that is being  probed and exploring the non-Gaussian profile of scalar/mixed/tensor interactions will enhance the characterization of the extra particle content in the EFT Lagrangian. We leave this to future work.

\noi As for the $s_0<0< s_2$ scenario, which corresponds to a blue(red)-tilted scalar(tensor) power spectrum, the growth of the scalar spectrum on small scales is reduced upon demanding the absence of gradient instabilities and complying with the bound on the tensor-to-scalar ratio $r<0.056$. 

\begin{figure}[H]
	\centering
	
	\includegraphics[width=0.45\linewidth]{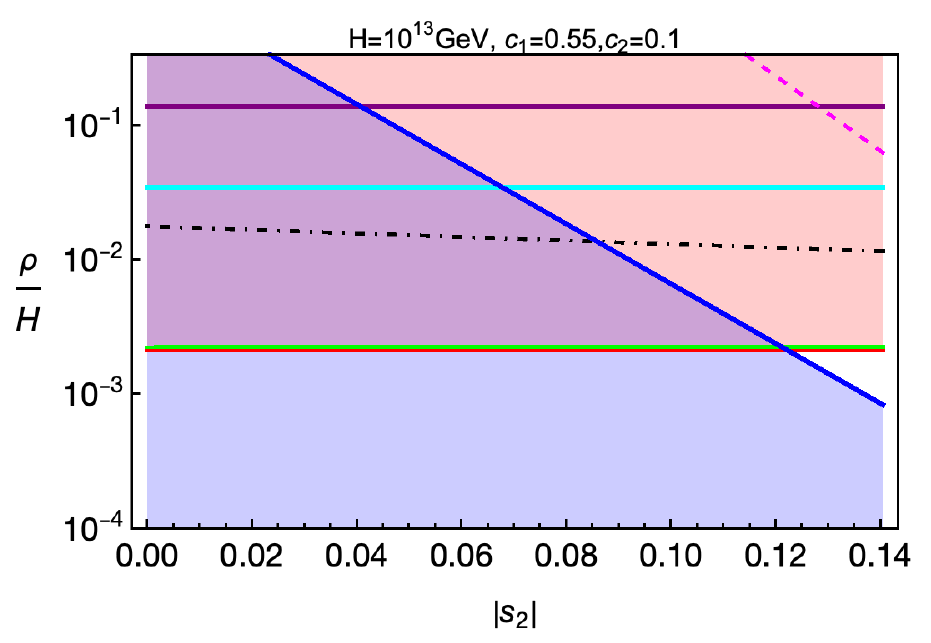} \quad \includegraphics[width=0.45\linewidth]{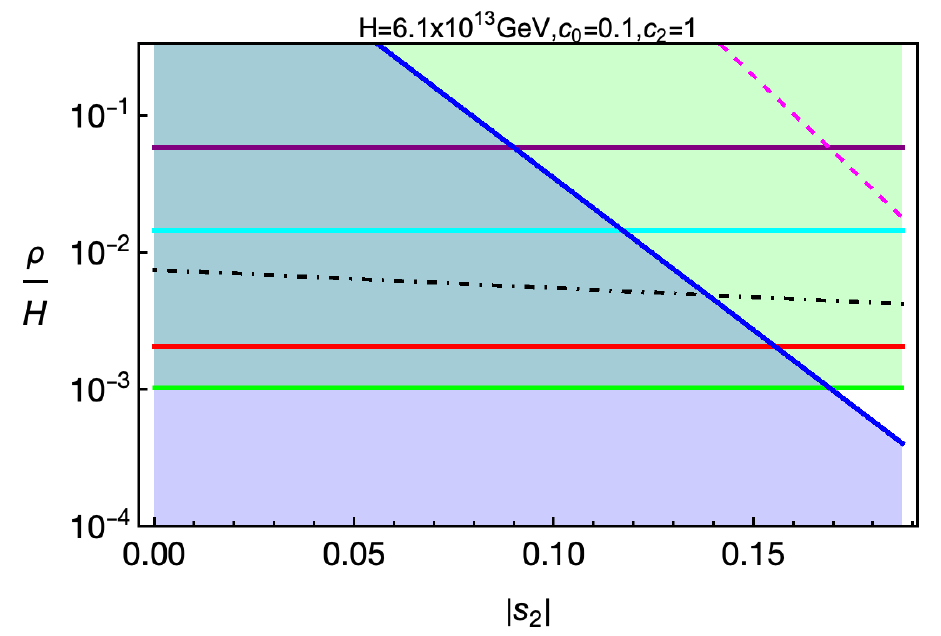}   \\ [\baselineskip]
	
	\includegraphics[width=0.45\linewidth]{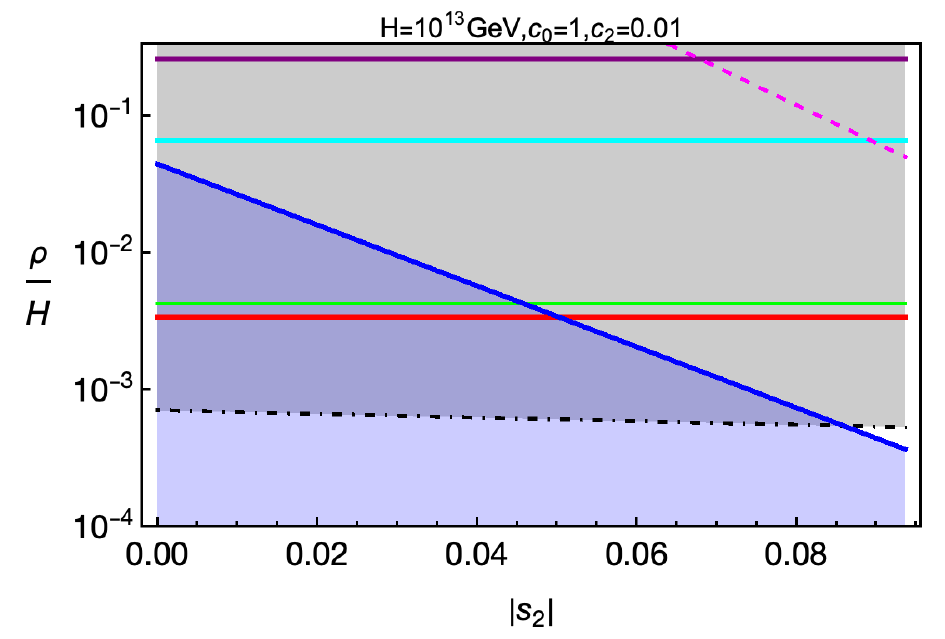} \quad \includegraphics[width=0.45\linewidth]{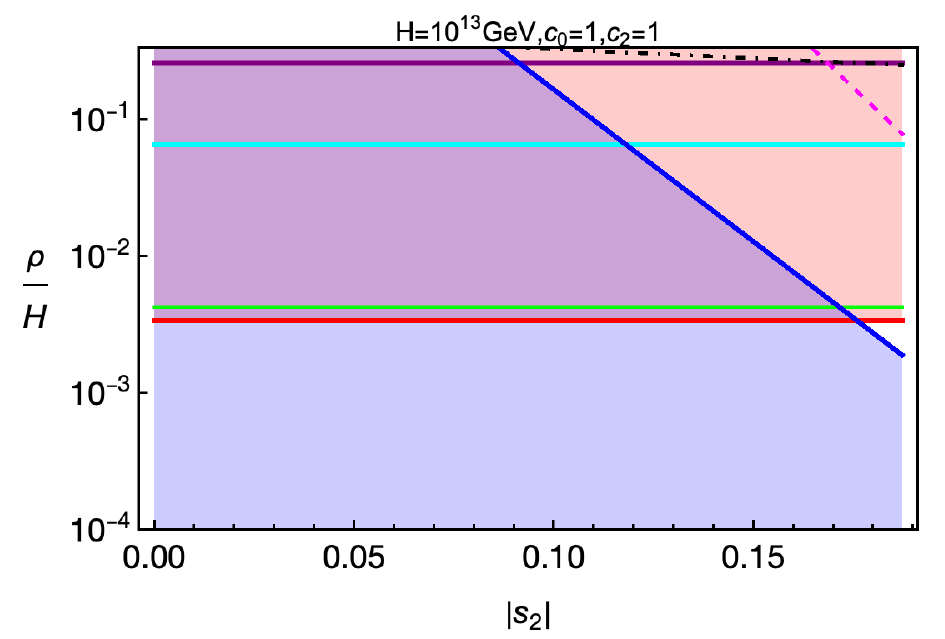}
	
	\caption{ The $(\rho/H, |s_2|)$ parameter space for samples from cases (1.a) and (2.a) which can be marginally surveyed by LISA. Conventions for colors and line codes are as in Fig.~\ref{fig:case1a}.}
	
	\label{fig:marginal}
	
\end{figure}

\begin{table}[t]
\centering
 \begin{tabular}{| c | c |} 
 \hline
 Case (1.a), Fig.~\ref{fig:case1a} & Case (2.a), Fig.~\ref{fig:case2a}  \\[0.5ex] 
 \hline\hline
  (b)$\{H=6.1 \times 10^{13}\, \mbox{GeV}, \, c_1=0.85, \, c_{2i}=1\}$ & (b) $\{H=6.1\times 10^{13}\, \mbox{GeV},\,  c_0=1, \, c_{2i}=1\}$ \\
 \hline 
 (c) $\{H=10^{13}\, \mbox{GeV},\, c_1=0.85, \, c_{2i}=0.1\}$ & (a) $\{H=10^{13}\, \mbox{GeV}, \, c_0=1, \, c_{2i}=0.1\}$
 \\
 \hline
\end{tabular}
\caption{\label{tab:similar_cases}  Configurations that select a similar portion of parameter space are listed on the same row of the table;  all the samples are characterized by the choice  $m/H=0.54$.}
\end{table}
\noi These limits turn out to be stronger than those that arise from PBH and UCMH. For completeness, we also list in Table \ref{tab:marginalconf} and plot in Fig. \ref{fig:marginal} the configurations whose parameter space can only be marginally surveyed by LISA.

\begin{table}[t]
	
	\centering
	
	\begin{tabular}{| c | c |} 
		
		\hline
		
		Case (1.a) & Case (2.a) \\[0.5ex] 
		
		\hline\hline
		
		$H=10^{13}\, \mbox{GeV},\, c_1=0.55, \, c_{2i}=0.1$ & $H=10^{13}\, \mbox{GeV},\, c_0=1, \, c_{2i}=0.01$\\
		
		\hline 
		
		& $H=10^{13}\, \mbox{GeV},\, c_0=1, \, c_{2i}=1$
		
		\\
		
		\hline
		
		& $H=6.1\times10^{13}\, \mbox{GeV},\, c_0=0.1, \, c_{2i}=1$
		
		\\
		
		\hline
		
	\end{tabular}
	
	\caption{\label{tab:marginalconf}: {Configurations for which LISA can only marginally survey the parameter space.}}
	
\end{table}

\section{Conclusions}
\label{sec:conclusions}

\noi In this work we have explored the possibility of a GW signal at interferometer scales due to the presence of extra fields during inflation. The existence of content beyond the minimal single-field scenario is well-motivated from the top-down perspective \cite{Baumann:2014nda}. Within this set, are there compelling models supporting in particular a tensor signal detectable at small scales? One interesting example is provided by so-called axion-inflation models \cite{Pajer:2013fsa}. The appeal of such set-ups lies in their ability to solve the $\eta$-problem: an approximate shift symmetry, for example in the ``natural inflation'' model of \cite{Freese:1990rb}, protect the inflaton mass from large quantum corrections. Extensions of the set-up in \cite{Freese:1990rb} have been motivated by the need to accommodate a sub-Planckian\footnote{The reasons \cite{Pajer:2013fsa} for why this is desirable are manifold: (i) the expectation that all global symmetries (including the aforementioned shift symmetry) are broken at the Planck scale; (ii) the near-absence of string theory constructions accommodating axions with a (super-)Planckian decay constant. One might point also out that a similar requirement emerges in the context of the swampland program (see e.g. \cite{Ooguri:2018wrx}).} axion decay constant $f$. Among the most interesting proposals to emerge from these efforts are those non-minimally coupling the axion-inflaton with gauge fields without losing the naturalness of the original proposal.  The information of interest to the present discussion is that the  coupling in such theories is typically implemented through a  Chern-Simons (CS) term. As a result of the CS coupling, the GW spectrum can be blue or exhibit bump-like features that peak at small scales \cite{Anber:2009ua,Barnaby:2010vf,Adshead:2012kp,Barnaby:2012xt,Maleknejad:2012fw,Dimastrogiovanni:2016fuu,Garcia-Bellido:2016dkw,Thorne:2017jft,Domcke:2018rvv}. \\
\noi Having identified a class of models that delivers a signal detectable by LIGO and/or, in the near future, the LISA interferometer, it is natural to ask whether several more multi-field set-ups, sharing this very same property, await discovery. One proven way to scan all that is possible, at least from the late-time signatures perspective, is to employ an effective theory approach. We have done so by adopting the approach of \cite{Bordin:2018pca} and focusing on the phenomenology corresponding to the presence of an extra spin-2 field. The key to a sufficiently large signal at small scales is choosing an appropriate time-dependence for the sound speed of the helicity-2 mode, $c_2(t)$. The existence of a time-dependence may be interpreted as due to a departure, in field space, from the adiabatic trajectory \cite{Achucarro:2012sm}. We have shown that a small and constant $s_2\equiv \dot{c}_2/(H c_2)$ corresponds to a signal to which LISA would be sensitive and that, at the same time, cannot be ruled out by LIGO/VIRGO. It would be interesting to explore other possibilities for the functional form of $c_2$; we leave this to future work. It is important to stress that a considerable region of parameter space has not been ruled out by existing data, but by the requirement \cite{Bordin:2018pca} that the dynamics does not run into a gradient instability. This goes to show that it is the interplay between model building \& observational requirements to act as the most powerful filter towards a viable cosmology.\\   

\noi The potential to detect a primordial signal must be confronted with our ability to: (i) distinguish it from astrophysical sources; (ii) identify signatures that are specific to certain (classes of) inflationary models. To address such issues one ought to characterize as much as possible the signal at small scales, and also consider cross-correlations with other cosmological probes. Given our results on the power spectrum, it is natural to think of non-Gaussianities as the next logical step. It has recently been shown that crucial information on the strength of primordial interactions is at least indirectly accessible at small scales \cite{Dimastrogiovanni:2019bfl}, and even once propagation effects are taken into account \cite{Bartolo:2019oiq}. A necessary ingredient to access the information via the quadrupolar anisotropies of \cite{Dimastrogiovanni:2019bfl} is a non-zero, and ideally large, component of the tensor bispectrum in the squeezed configuration. The presence of light extra fields in the inflationary set-up we have been studying supports precisely such a scenario. Furthermore, the contribution in the squeezed configuration mediated by the extra content will break consistency relations and therefore deliver a signal that is immediately physical. We shall elaborate more on the subject in upcoming work \cite{Iacconi:2020yxn}.

\acknowledgments
It is a  pleasure to thank Gianmassimo Tasinato for collaboration at the early stages of this project.
LI would like to thank Enrico Barausse, Germano Nardini, Antoine Petiteau, and Angelo Ricciardone for very useful input on LISA sensitivity, as well as Laura Nuttall, Ian Harry and Andrew Matas for useful discussions about LIGO bound. HA, MF, LI, and DW are supported in part by STFC grants ST/N000668/1 and ST/R505018/1.

\bibliography{November_6} 

\bibliographystyle{JHEP}

\end{document}